# Introducing antiferromagnetic ordering on the surface states of Bi$_2$Se$_3$ topological insulator by Europium doping

Sumana Paul[1][¶], Moumita Das,[2,3] Sujoy Datta[4], Raja Chakraborty[3], Prabhat Mandal[2] and P. K. Giri[1,5]∗

[1]Centre for Nanotechnology, Indian Institute of Technology Guwahati, Guwahati 781039, India

[2]Saha Institute of Nuclear Physics, HBNI, 1/AF Bidhannagar, Kolkata 700064, India

[3]School of Physical Sciences, Indian Association for the Cultivation of Science, Jadavpur, Kolkata-700032, India

[4]Kadihati KNM High School, Ganti, Kolkata 700135, India

[5]Department of Physics, Indian Institute of Technology Guwahati, Guwahati 781039, India

¶ Present Address: CSIR-Central Glass and Ceramic Research Institute, 196 Raja S. C. Mullick Road, Kolkata 700 032, India

## ABSTRACT

Topological insulators (TIs) are materials with an insulating bulk characterized by a gapped band structure, along with gapless metallic surface states having a Dirac cone with a helical spin structure in momentum space. The helical spin–momentum locking of the surface states arises from intrinsic spin-orbit coupling (SOC) and provides topological protection to the surface states against scattering from external perturbations, like defects and non-magnetic impurities. Breaking the topological protection of surface states of topological insulators is an essential prerequisite for exploring their applications. Rare-earth ions typically exhibit larger magnetic moments than transition-metal ions and thus promise the opening of a wider exchange gap in the Dirac surface states of topological insulators. Bi$_2$Se$_3$ is an interesting material; on the one hand, it has semiconducting properties when it is thin sheets; on the other hand, it's a topological insulator when the structure has a minimum of six quintuple layers (QLs), with diverse applications in photothermal, thermoelectric, and optical properties. Here, we have developed a controlled colloidal synthesis with low temperature and a cost-effective process for the synthesis of undoped and Eu-doped 2D layered $Bi_2Se_3$ nanosheets. Scanning tunneling spectroscopy measurements demonstrated a correlation between the shift of Dirac point position and the dopant content, which has been theoretically established by the band structure calculation. At low temperatures below 10K, for the 10% Eu doped sample, magnetic data suggests an antiferromagnetic ordering in the sample, which may be the contribution of the mixed valance state of Europium. Our results support that antiferromagnetic exchange interaction can exist in topological surface states in rare earth Eu-doped $Bi_2Se_3$, which can open a new window to novel quantum phenomena.

∗Corresponding author; email: giri@iitg.ac.in

## 1. Introduction

The family of 2D crystals is emerging as new stars in the study of electronic, optoelectronic, and magnetic properties due to their high aspect ratios, tunable bandgaps, and diverse optical properties.[1-3] More often, a few layers of these crystals possess properties that are strikingly different from those of their 3D counterpart: for example, graphene is a zero-gap semiconductor, whereas graphite is a semimetal with a band overlap; there are several transition metal dichalcogenides (TMDC) in 2H phase with direct-bandgap semiconductors, whereas in their bulk form, the bandgap is indirect.[4-5] These materials with different properties have led to the creation of unique heterostructures and allowed the investigation of new physical insights, as well as the development of new devices.

Recently, Topological Insulator (TI), emerging as a new quantum state of matter, has drawn enormous interest in condensed matter physics due to their novel physical phenomenon of insulating or semiconducting properties in bulk and metallic Dirac states at the surface.[6-8] TIs are materials that have a bulk electronic band gap and metallic surface states for charges and spins, which are protected by time-reversal symmetry. The unique transport properties and high mobility are attributed to the prevention of backscattering on the topological surface states, which makes them suitable for advanced electronic devices. It is interesting to note that TI-based devices show several fascinating applications, such as quantum computing, spintronics, thermoelectrics, photonics, and high-speed optoelectronics.[9-13]

Among the various TIs, Bismuth selenide ($Bi_2Se_3$) attracts the growing attention of a broad research community as a layered semiconductor. Initially, $Bi_2Se_3$ was extensively studied as a thermoelectric material.[14] However, in recent years, a large number of reports have emphasized the deep understanding of the quantum-mechanical properties of this material and opened it to new areas, such as spintronics. A strong spin−orbit coupling makes the surface states of a topological insulator invert their orbital character to fulfill matching conditions at the interface with vacuum or other materials having weak spin−orbit coupling. This leads to the formation of two crossing surface state bands between bulk valence and conduction bands, giving rise to the appearance of Dirac cones and metallic conduction.[15-16] $Bi_2Se_3$ has been a well-studied material for its special electronic property of simple surface state structure with a single Dirac cone and a relatively large bandgap value of ~ 0.35 eV in bulk.[17] $Bi_2Se_3$ possesses unique physical phenomena like quantum oscillation,[18] weak antilocalization[19], which are interesting for future applications. There are several interesting properties it has like: high surface mobility (~$10^4$ $cm^2V^{-1}s^{-1}$),[20] superior electrical conductivities,[21] and excellent photoconductivity[22] properties. The small intrinsic bandgap helps $Bi_2Se_3$ to absorb more light over a broad wavelength range, while the topological structure makes them robust and stable in ambient and harsh environmental conditions. In addition, $Bi_2Se_3$ shows exciting optoelectronic properties, such as size and thickness-dependent band gap, the fast dynamical response of electrons under optical excitation, polarization-sensitive photocurrent, and many more.[23-25]

Several growth strategies have been adopted to prepare high crystalline $Bi_2Se_3$ flakes/ sheets like molecular beam epitaxy (MBE),[26] mechanical exfoliations,[27] chemical vapor deposition (CVD),[28] liquid phase exfoliation[29], etc. However, yield in CVD or MBE-like methods is very low in comparison to solution-based processes and also needs designated substrates and high temperatures, which is not very simple and cost-effective. Solution-processable synthesis procedures have already been shown to be a versatile tool for producing large NSs with nanoscale thicknesses and size tunability. Also, morphology variation and doping are possible in this synthesis procedure.[30-31] Also to reduce the synthesis temperature and the cost, solution-processable synthesis techniques are an emerging field for material synthesis at low temperatures with excellent properties and high crystallinity.

A majority of the studies on the doping of TIs have focused on enhancing their magnetic properties.[32-34] The topological protection can break down under certain circumstances, e.g., in the presence of magnetic impurities.[35] A magnetic impurity can cause spin-flips and allow backscattering of the surface states, hence leading to enhanced surface resistivity. When the exchange interaction among the magnetic impurities is sufficiently strong, a long-range ferromagnetic (FM) ordering may arise, and a finite gap is opened at the Dirac point. Such a gap prevents the scattering of surface states to the bulk continuum and hence gives rise to a number of interesting phenomena, such as the quantum anomalous Hall effect (QAHE), topological magneto-electric effect, etc.[36-37] Enormous efforts have been made regarding the intentional magnetic doping of $Bi_2Se_3$ since the discovery of three-dimensional topological insulators (TIs). Ferromagnetic ordering has been reported in Cr, Fe, and Cu doped $Bi_2Se_3$ thin films,[38-40] Cr and V doped $Sb_2Te_3$,[41] Mn doped $Bi_2Te_3$ single crystals, etc.[42] Rare earth (RE) atoms often have large atomic magnetic moments and they are comparable in size to the Bi atom creating a large Dirac gap. The large magnetic moment of the RE elements originates from the unpaired 4*f* electrons. The occurrence of paramagnetism down to 2 K in MBE-grown Dy-doped $Bi_2Te_3$ thin films has already been studied.[43] A paramagnetic phase due to the low doping concentration of only 1 at% has been reported in the Gd-doped $Bi_2Se_3$.[44] In a series of Sm-doped $Bi_2Se_3$ MTIs,[45] ferromagnetisms up to about 52 K and a suppressed bulk electron carrier concentration has been reported. Apart from ferromagnetic ordering, antiferromagnetic ordering has also been acting as a distinct flavor of magnetic order which has recently drawn significant attention with respect to various spintronic applications. Among them, MnBi$_2$Te$_4$ and high-quality single-layer FeSe thin film have already been reported as antiferromagnetic topological insulators.[46-47] A new two dimensional magnetic topological insulator EuCd$_2$Bi$_2$ has also been emerged with stable crystal structure and giant non-trivial gap, reaching as much as 750 meV. It has been studied that the magnetic TI phase survives under various magnetic configurations, including ferromagnetic and antiferromagnetic coupling with both in-plane and out-of-plane directions, in 2D EuCd$_2$Bi$_2$ quintuple layers (QLs).[48] Niu *et al.* has predicted that XMnY (X = Sr and Ba, Y = Sn and Pb) quintuple layers (QLs) are a family of long-awaited intrinsic nonsymmorphic 2D antiferromagnetic TIs.[49] It is well-known that RE chalcogenides can exhibit antiferromagnetic ordering. So, it will be promising to take advantage of the large RE moments by doping RE elements in TIs to enhance antiferromagnetic ordering which will be very promising platform to achieve topologically complex insulating materials for antiferromagnetic spintronics.

In light of the above, the present work aims at exploring structural, optical, and magnetic properties of as-prepared undoped and rare earth Eu doped Bi$_2$Se$_3$ NSs, synthesized using a wet chemical method. We show in the following that antiferromagnetic ordering appears in the Eu-doped samples. The details about the crystal structure and morphology of the Eu-doped Bi$_2$Se$_3$ nanosheets were characterized by XRD and TEM analyses. The doping effect is confirmed by the shift in the Raman peaks as well as elemental mapping. The band edge absorption and change in the band gaps of the samples were investigated systematically by UV-vis-NIR spectroscopy. The room temperature scanning tunneling spectroscopy data and density of states calculations suggest an enhancement in the band gap of the doped samples. Rare earth-doped TIs have been studied very little in the literature. The results demonstrated in this report signify that Eu doping induces antiferromagnetic ordering into the topological microsheets. This would provide a new direction for the next generation of TI-based spintronic devices.

## 2. A. Experimental section

**Materials:** All chemicals were used as-purchased directly without any purification. Bismuth nitrate pentahydrate (Bi(NO$_3$)$_3$·5H$_2$O, >99.9%), Sodium selenite (NaSeO$_3$, >99%), 90%], Europium nitrate pentahydrate (Eu(NO$_3$)$_3$·5H$_2$O, >99.9%), Polyvinylpyrrolidone (PVP) (average MW 40,000) were purchased from Sigma-Aldrich. Ethylene glycol (EG), toluene, isopropanol, ethanol, and other solvents were purchased from Merck, India. All synthetic experiments were conducted using a Schlenk line under a dry N$_2$ atmosphere.

**Synthesis of Bi$_2$Se$_3$ and Eu doped Bi$_2$Se$_3$ ((Eu$_x$Bi$_{1-x}$)$_2$Se$_3$) microsheets (MSs):** In a typical synthesis procedure, 0.2 mM of Bi(NO$_3$)$_3$, 5 H$_2$O, 0.3 mM of NaSeO$_3$ and 2 mM of PVP and 10 ml of EG were loaded in a three neck round bottom flask fitted with a condenser. The mixture was then evacuated at ambient temperature for 10 min and at 45°C for another 20 min. Subsequently, the N$_2$ gas was purged into the solution and the temperature was increased rapidly to 190°C and the reaction was continued for 2.5 hr at this temperature. After completion of the reaction, the solution was cooled down to room temperature naturally. The obtained sample was washed with acetone and isopropanol as a non-solvent, followed by centrifugation at 10000 rpm for 3 min. This process was repeated 3-4 times, and finally, the sample was dispersed in isopropanol for further use.

The doped (Eu$_x$Bi$_{1-x}$)$_2$Se$_3$ samples were synthesized under the same reaction conditions by adding (Eu(NO$_3$)$_3$·5H$_2$O in the required stoichiometric ratio in EG, where x = 0.025, 0.05, 0.075, and 0.1.

**Characterization:** The X-ray powder diffraction (XRD) method was used to determine the crystalline phases of the as-synthesized samples by using a Bruker AXS D8SWAX diffractometer with Cu Kα radiation (λ = 1.5418 Å), employing with a scan rate of 0.5° S$^{-1}$ in the 2θ range from 5° to 60°. For XRD measurements, the samples dispersed in isopropanol were drop-casted over an amorphous glass substrate until a thin layer visible to the naked eye was formed. The transmission electron microscopy (TEM) images, including high-resolution TEM images and energy dispersive spectra (EDS) for elemental mapping of the as-synthesized samples, were taken by an Ultra-high-resolution field emission gun transmission electron microscope (UHR-FEG TEM, JEM-2100F, JEOL, Japan) operating at 200 kV. For the observation of the samples in TEM, the samples dissolved in isopropanol were drop-casted on a carbon-coated copper grid. Raman scattering measurements were performed in a high-resolution Raman spectrometer (LabRam HR800, Jobin Yvon) with laser excitation 633 nm. A 100X objective lens focused the laser beam with a spot size of ∼1 μm in diameter, and the acquisition time was 15 s for collecting the Raman spectrum of the different samples. The room temperature optical absorption of the samples dissolved in toluene was measured by a Varian Cary 5000 UV-VIS-NIR spectrometer. The thickness of the as-synthesized $Bi_2Se_3$ MSs was examined using a Bruker Innova atomic force microscope (AFM) operating in the tapping mode. A small piece of rectangular shape sample was cut from the as-prepared sample for the magnetization (M) measurements. The temperature and magnetic field dependence of magnetization measurements have been carried out using a superconducting quantum interference device-vibrating sample magnetometer (SQUID-VSM, Quantum Design). The sample space was degaussed every time before loading the sample in the magnetometer to avoid any stray magnetic field originating from the superconducting magnet. The magnetization, as a function of temperature and magnetic field for the empty sample holder, was recorded before performing the magnetic measurements to negate its diamagnetic contribution towards the actual magnetization value of the as-synthesized TIs. Zero-field-cooled (ZFC) data was collected after cooling the sample to the lowest temperature in the presence of $H = 0$ and taking the data during the heating cycle under the applied magnetic field H=10kOe. Field cool (FC) was taken during the cooling cycle in the presence of an applied magnetic field H=10 kOe. *M-H* hysteresis loops were collected after cooling the sample to a specific temperature under ZFC conditions.

To form an ultrathin film of the materials on cleaned arsenic-doped silicon substrates having a resistivity of 5-10 mΩ-cm, the spin coating was carried out at 3000 rpm for 30 s, followed by an annealing step at 60 °C for 10 min. The silicon substrate was cleaned with DI water, acetone, and alcohol and treated with 4% dilute hydrofluoric (HF) to remove all traces of organic contaminants and native oxide layers, followed by ultraviolet-ozone treatment.

The materials were probed with a Nanosurf Easyscan2 scanning tunneling microscope (STM) to record scanning tunneling spectroscopy (STS) spectra under ambient conditions. During the approach of the mechanically cut extremely sharp tips of Pt/Ir (80%:20%), a current of 500 pA was set at a bias of 1.0 V.

**B. Theoretical Calculations:** For a better understanding of the underlying physics of the observed properties, density functional theoretical (DFT) calculations using the Vienna Ab-initio Simulation Package (VASP) package were used.[50] Vaspkit was utilized for the post-processing of results.[51] Perdew-Burke-Ernzerhof (PBE) formulation of generalized gradient approximated (GGA) exchange-correlation (xc) functional applied for Projected augmented wave (PAW) basis set was employed throughout.[52] The doped structure was made by randomly replacing Bi atoms with Eu atoms in a 2x2x2 supercell containing 8 formula units. Both of the primitive and doped structures were optimized through ionic and volume relaxations. The structural parameters are provided in the Appendix. We set the force threshold for ionic relaxation as $10^{-3}$ eV/Å. and the threshold for pressure as and $10^{-2}$ Kbar/cell. For convergence of total energy and charge densities, the criteria were set as $10^{-8}$ eV. The kinetic-energy cut-off was 280eV for all calculations. Reciprocal-space grids of dimensions 8x8x8 were used for primitive structures and 2x2x2 for the doped ones. For better convergence of the self-consistency loop, an additional support grid was incorporated for Fast-Fourior-Transformation at the time of augmentation of charges. Eu belongs to the lanthanide series, where f-electrons play a role in its electronic structural properties. For atoms with f-electrons, the aspherical contribution to Kohn-Sham potential and total energy within PAW spheres, originating from gradient correction for charge density, is important and, hence, applied.

## 3. Results and discussion

### 1. Structural and Optical Studies

**Structural and Morphological Analysis:**

Wet-chemical synthesis procedures are a versatile tool for producing microsheets (MSs) of large sizes and nanoscale thickness. Here, we have synthesized two-dimensional (2D) MSs of $Bi_2Se_3$ with large lateral sizes by using an EG-based method, where PVP has been introduced as a stabilizer. To demonstrate the prospective effect of doping, we have synthesized solution-processable $Bi_2Se_3$ MSs and Eu-doped $Bi_2Se_3$ MSs. $Bi_2Se_3$ is an anisotropic layered material that consists of quintuple layers (1 QL, ≈9.55 Å) with consecutive five atomic layers Se1-Bi-Se2-Bi-Se1 stacked along crystallographic c-axis by a weak vdWs interaction as depicted in Figure 1a.[53] The pink-shaded region represents the primitive unit cell. The c-axis of the conventional cell is the diagonal of the primitive one. X-ray diffraction (XRD) plot, as shown in Figure 1c, was employed to identify the crystal structure and the phase purity of the samples. Its resemblance with the simulated XRD pattern of Figure 1b signifies the purity. The crystal structure of the as-grown samples shows rhombohedral crystal geometry with a space group belonging to (R$\bar{3}$m) (JCPDS No. 33-0214) with no other impurity phases. While the simulated XRD pattern shows peaks related to different planes, it has been noticed from the measured XRD pattern that only {003} family of diffraction peaks are prominent, indicating a preferred growth direction along the c-axis. The most prominent measured peak originates from the (006) plane at 2θ = 18.2°, which agrees well with the simulated value of 2θ = 17.25°. Here, the QL layers are attached to each other by weak van der Waals interaction. The XRD patterns of the doped samples follow similar patterns to those of the pure ones, with no other impurity phases related to Eu. The closer view of the (006) plane shows a shift toward the higher 2θ value, as displayed in the inset of Figure 1d. The peak position shifting with the doping concentration follows a nonlinear relationship, as indicated in Figure 1d, which has been reported in many doped systems. This shifting towards a higher 2θ value might be due to the smaller ionic radius of $Eu^{+3}$ (ionic radius=1.06Å) than that of $Bi^{+3}$ (ionic radius=1.17Å). This fact signifies the successful substitution of Eu into the $Bi_2Se_3$ matrix.[54] The morphology and the thickness of the wet chemically grown MSs were analyzed by atomic force microscope (AFM). The AFM topography of a single MS shows (Figure 1e) hexagonal morphology with a lateral dimension of ~800 nm. The height profile of ~7.2 nm measured from the AFM image (the inset of Figure 1e) indicates the ultrathin nature of the sample (~7-8 layers). The AFM topography of the doped sample shows a similar morphology and thickness slightly decreased (~ 6 nm) with Eu doping, as depicted in Figure 1f.

Transmission electron microscope (TEM) images were studied to investigate further the morphology, crystal structure, and growth orientation of the as-synthesized MSs. The overall morphology shows hexagonal MSs, as displayed in Figure 2a. Figure 2b shows a closer view of a single hexagonal MS. The lateral size of the MSs varies between 0.5-3 μm. The high-resolution TEM (HRTEM) image is shown in Figure 2c, and the inset shows a Fast Fourier Transform (FFT) pattern as obtained from the yellow box in the same figure. The FFT pattern reveals a sixfold symmetry [0001] axis of the stacked-layered structure of $Bi_2Se_3$ along the c-axis. The reconstructed HRTEM image (Figure 2d) shows a *d* spacing of 0.21 nm, corresponding to the group of planes (11-20). We have further verified the presence of Bi and Se by high-resolution energy-dispersive X-ray (EDS) mapping analyses (Figure 2e). The Bi and Se elements (Figures 2f and 2g) are homogeneously distributed over a single hexagonal nanosheet with an atomic ratio of Bi: Se ≈ 0.79 (Figure 2h), which is higher than the expected ratio of 0.67. This implies a Se deficiency (vacancy) in the $Bi_2Se_3$ crystal.

The bright field TEM image of 5% Eu doped $Bi_2Se_3$ is shown in the Electronic Supplementary Information (ESI) in Figure S1a, which shows a hexagonal sheet-like morphology, and the lateral size ranges between 0.5-3.0 μm. A single hexagonal sheet is shown in Figure S1b. The HRTEM image (Figure S1c) confirms the well-crystalline nature of the doped samples. The FFT pattern, as obtained from the green box in Figure S1c, shows the (11-20) group of planes. The lattice spacing in HRTEM (Figure S1d) is ~0.209 nm, slightly lower than that of pure $Bi_2Se_3$ MSs, as expected due to the Eu doping. The compositions of doped MSs were determined by energy dispersive X-ray spectroscopy (EDS) and elemental mapping (Figure S1e-i), which revealed successful Eu doing in $Bi_2Se_3$ MSs.

Figure S2a shows a large area bright field TEM image of 10% Eu doped $Bi_2Se_3$ MSs, which displays hexagonal MSs. The Eu doping concentration does not affect the lateral size and morphology of the system. Figure S2b shows a bright field TEM image of a single MS, and the HRTEM (Figure S2c) shows the crystalline nature of the sample. The FFT pattern shows the (11-20) set of planes (as depicted in the inset of Figure S2c), and the corresponding *d* lattice spacing is 0.205 nm (Figure S2d), which is slightly smaller than that of the pure one due to the doping effect. The compositions of doped MSs were determined by EDX and elemental mapping, which revealed the incorporation of Eu in $Bi_2Se_3$ MSs. The elemental mapping over a few MSs shows a homogeneous distribution of Bi, Se, and the dopant Eu, as displayed in Figure S2(e-i).

The core-level XPS allows to probe the oxidation state and the chemical environment of Eu-doped $Bi_2Se_3$ MSs. The high-resolution XPS spectra of Bi 4f, Se 3d, and Eu 3d in $(Eu_{0.1}Bi_{0.9})_2Se_3$ MSs can be fitted properly using Shirley baseline and Lorentzian peak fitting. The two peaks Bi $4f_{7/2}$ and $4f_{5/2}$ centered at ∼157.6 and ∼163.0 eV (Figure S3a), respectively, with a separation of ∼5.4 eV is consistent with the previous report (Figure S3a). These peaks are indicative of the Bi(III) valence state in $Bi_2Se_3$. Similarly, the high-resolution spectrum of Se 3d (Figure S3b) is fitted with two peaks: Se $3d_{5/2}$ and Se $3d_{3/2}$ peaks with binding energies centered at 51.8 and 52.6 eV, which are characteristic of the Se(II) valence state (Figure S3b). Next, the high-resolution XPS spectrum of Eu 3d (Figure S3c) is investigated to ascertain the valance state of europium. Here, Eu 3d XPS peaks show four peaks centered at $Eu^{3+}$ $3d_{5/2}$ (1133.6 eV) and $Eu^{3+}$ $3d_{3/2}$ (1162.7 eV) and $Eu^{2+}$ $3d_{5/2}$ (1124.6 eV) and $Eu^{2+}$ $3d_{3/2}$ (1154.5 eV). Therefore, a mixed-valence state, $Eu^{3+}/Eu^{2+}$, is observed in the Eu-3d XPS pattern.

Raman spectroscopy measurements were carried out to measure the Raman shift due to the Eu doping, which affected the lattice structure. Raman spectra of the as-synthesized MSs were obtained at a laser excitation of 633 nm in the wavenumber range 60−200 $cm^{-1}$ at room temperature. Three prominent Raman peaks were observed at 72.5, 132.3, and 174.9 $cm^{-1}$, assigned to out-of-plane $A_{1g}^1$, in-plane $E_g^2$, and out-of-plane $A_{1g}^2$, respectively, as represented in Figure 3. The Raman peaks of 10% Eu-doped samples were observed at 71.9, 130.7, and 173.5 $cm^{-1}$. The Raman shift of Eu-doped $Bi_2Se_3$ shows a red shift as compared to the undoped one. The peak intensities were observed to be decreasing simultaneously, as observed in the case of other doped topological insulators. According to the previous studies on phonon dynamics of TIs, the shift may arise due to an increasing lattice strain. There may be several reasons behind size-induced phonon redshift, like surface disorder, surface stress, and phonon quantum confinement.[55] The phonon confinement model suggested that strong phonon damping happens with decreasing solid size. In this work, Eu doping induces a strain enhancement in the material, and the frequencies of the two optical phonons, $E_g^2$ and $A_{1g}^2$, are red-shifted. Compared with the Bi–Se bond, the Eu–Se bond is shorter (Eu has a lower atomic radius than that of Bi), the QLs become more compact, extra compression strain occurs, and the enhanced van der Waals interaction, which results in a higher phonon vibration frequency.[56]

**Optical properties:**

A suspension of the bare $Bi_2Se_3$ and Eu-doped $Bi_2Se_3$ MSs in isopropanol was investigated by UV-vis-NIR spectroscopy in the spectral range 200-1400 nm, as shown in Figure 4a. Interestingly, both the undoped and doped samples exhibit very broadband absorption with high absorbance, and the absorbance is found to increase systematically with increasing doping concentration of Eu. The bandgap absorption edge of undoped $Bi_2Se_3$ was found to be around ~600 nm, which is red-shifted upon increasing Eu doping concentration, and for the 10% doped sample, it is at ~640 nm. We have also calculated the absorption coefficient of the thin nanosheets in our previously reported study on Eu-doped $Bi_2Se_3$ MSs.[54] The absorption coefficients of the undoped sample, as well as the doped sample, were found to be as high as the conventional bulk semiconductors. Generally, the bandgap value of $Bi_2Se_3$ single crystals varies between 0.22 to 0.33 eV[57], and that of thin films varies between 0.1 to 2.0 eV.[57-59] The optical band gap ($E_g$) of the as-synthesized microsheets was calculated by using the Tauc plot: $(\alpha h\nu)^2$ vs. $h\nu$ based on the following equation for direct bandgap semiconductors:[60]

$$(\alpha h\nu)^2 = A(h\nu - E_g) \qquad (1)$$

where α is the optical absorption coefficient, hν is the incident photon energy, $E_g$ is the bandgap, and A is a constant. $Bi_2Se_3$ is a direct bandgap material, and the calculated optical band gap value of undoped $Bi_2Se_3$ is found to be 1.29 eV (Figure 4b), which is higher than the bulk $Bi_2Se_3$ rhombohedral crystal structure. From the AFM study, we observed that the average thickness of the MSs is ~7 nm, which is relatively low. Low dimension can lead to a quantum confinement effect in different directions, and the surface may be responsible for the larger band gap value. $Bi_2Se_3$ MSs have strong absorption in the visible region, which originates from the Se vacancies.[61] Imperfections due to the Se vacancy affect the optical properties by modifying charge carrier density as well as band structure. Surface capping agents can interact with the surface charges of the MSs and can change the surface charge density. It is very challenging to control Se vacancies in wet chemical synthesis, and it is quite apparent to get higher absorption in the visible region.

Doping in topological insulators is a promising approach for bandgap tuning. The bandgap value changes from 1.29 eV to 1.07 eV upon Eu incorporation into the $Bi_2Se_3$ matrix (inset of Figure 4b). Referring to the theoretically calculated DOS as presented in Figure 4c, it is evident that while $Eu^{3+}$ doping increases the bandgap, the bandgap closes for $Eu^{2+}$ doped systems. The impact of $Eu^{2+}$ and $Eu^{3+}$ doping in $Bi_2Se_3$ is elucidated through a bonding and anti-bonding analysis. Crystal orbital Hamilton population (COHP) plot is a powerful tool to identify the bonding and anti-bonding between different ions in a solid.[62] Traditionally (-COHP) is plotted to show the right side as bonding and left side as anti-bonding interaction (Figure 4d).[63] We used Lobster package to calculate the (-COHP).[64]

For pristine $Bi_2Se_3$, 11.4% anti-bonding interaction between Bi-Se pairs is present below Fermi energy. For doped systems, due to symmetry breaking, three different types of Bi ions emerge. The anti-bonding percentage for Bi-Se pairs ranges from 10.0% to 14.7% in $Eu^{2+}$-doped $Bi_2Se_3$, and from 10.9% to 12.2% in $Eu^{3+}$-doped $Bi_2Se_3$. At Fermi energy, there is no -COHP for the pristine and $Eu^{3+}$ doped systems, hence, bandgaps open up for these two. Whereas, in $Eu^{2+}$ system, a finite anti-bonding state is present for both Bi-Se and Eu-Se pairs. In the $Eu^{3+}$-doped system, the Eu-Se pair exhibits no anti-bonding state below E_F, whereas in the $Eu^{2+}$-doped $Bi_2Se_3$, the Eu-Se pair has a 17% anti-bonding state below $E_F$. So, the effect of doping of these two types of Eu ions in the electronic property are quite different in nature as supported by the experimental findings.

For these structures, each of Bi and Eu atoms resides at octahedral coordination environment having six Bi/Eu-Se bonds. Below Fermi energy no Eu-Se anti-bonding states were formed in $Eu^{3+}$ doped system, whereas for $Eu^{2+}$ system 17% Eu-Se anti-bonding states were formed. It is well known that in nature, the higher energy anti-bonding state hinders the stability, hence structures with less anti-bonding state at or below Fermi level is more preferred (e.g., hexagonal Te is more stable than cubic Te).[65] So using this analysis we can easily understand why the occurrence of $Eu^{3+}$ doping is more preferred while doping $Bi_2Se_3$.

From the XPS study, we also concluded that both $Eu^{2+}$ and $Eu^{3+}$ states are present in the system. Therefore, the observed decreasing trend of the bandgap is attributed to the presence of the mixed state in the synthesized samples. The reduction of band gap upon magnetic ion doping (Fe) has already been reported in the topological insulator $Bi_2Se_3$.[66] There may be several reasons for the reduction of band gap value upon doping. One of the main reasons is a large difference in the electronegativity of Bi (2.02) and Eu (1.2), which has been reported in the Tl-doped $Bi_2Se_3$.[67] Adding impurities (donor/acceptor) into semiconductors creates new energy levels near the conduction or the valence band edges. An increase in the dopant concentration will increase the density of these energy states, and as a result, a continuum of states can be formed.[68] Therefore, the bandgap decreases with increasing Eu concentration. Inset of Figure 4b shows the plot of bandgap variation with the increase of Eu doping concentration, and it follows a linear relationship between bandgap and doping concentration. This clearly demonstrates the successful doping into the $Bi_2Se_3$ system as compared to previously reported systems. Doping with Eu ions will increase the absorbance in the IR region. Maiti et al. have reported that absorbance in the IR region increases with the decreased thickness of the $Bi_2Se_3$ MSs.[69] Raman study of the samples also exhibits a redshift of the $A_{1g}^1$ peak and broadening of the Raman peaks, which are the typical characteristics of thickness decrease on doping.

**Scanning tunneling spectroscopy: density of states and band edges**

To further shed light on the effect of Eu doping, we proceeded to record d$I$/d$V$ spectra of the different materials in their ultrathin forms, which corresponded to the energy-dependent density of states (DOS) of the material at the point of measurement. At either side of the Fermi energy (EF), which is supposed to be aligned at 0 V, generally, DOS spectra provide the valence band (VB) and conduction band (CB) edges of the material in the form of the first peak. During the measurement, as the bias was applied to the STM tip, a positive tip voltage implies the withdrawal of electrons from the sample; thus, the first peak denotes the VB edge of the material. Similarly, a negative tip voltage implies the injection of electrons to the CB edge of the material. The d$I$/d$V$ spectra of all of the materials have been represented in Figure 5a. To start with the pure $Bi_2Se_3$ material, the V-shaped DOS and non-zero tunneling current are observed with the Dirac point at -0.19 eV. Non-zero DOS at the position of $E_F$ and zero DOS at an energy away from the Fermi energy imply the topological nature of $Bi_2Se_3$. However, the Dirac point is in the negative voltage, i.e., below the Fermi energy, which implies that the pure $Bi_2Se_3$ is basically n-type in nature. The doping of Eu in $Bi_2Se_3$ has a direct impact on the d$I$/d$V$ spectra of the material under consideration. Upon 2.5% doping with rare earth Eu, the Dirac point shifts to -0.13 eV, and it is further moved to -0.06 eV for 5% Eu doping. However, a further increase in Eu doping in $Bi_2Se_3$ leads to significant changes in the d$I$/d$V$ spectra. Interestingly, the trace of the Dirac point disappears upon further doping of Eu in the material, and DOS shows a small band gap in the range between 0.3 - 0.4 eV. In the case of 7.5% Eu doped $Bi_2Se_3$, the CB and VB positions are -0.12 eV and 0.29 eV, respectively, with a band gap of 0.41 eV. Whereas when Eu doping increased to 10%, the band gap was slightly reduced to 0.34 eV, and the corresponding CB and VB positions were at -0.10 eV and 0.24 eV, respectively.

A theoretical understanding is always helpful in comprehending the observed phenomena. In Figure 5, we present the band structure of pristine and $Eu^{3+}$ doped structures of different concentrations. As observed in Figure 5(b), the nature of the bandgap in $Bi_2Se_3$ is direct at the Brillouin Zone (Figure 5c) center ($\Gamma$). The gap is found as 0.61 eV, which becomes larger with $Eu^{3+}$ doping, 0.75 eV at 6.25%, and 0.78 eV at 12.5%. It clearly complements the experimental finding of the role of $Eu^{3+}$ doping in increasing the gap. Upon doping, the valence band becomes flatter, so it is not easy to determine whether it is a direct or indirect semiconductor, as depicted in Figures 5(d) and (e).

## 2. Magnetic Study:

The main panel of Figure 6(a) shows the magnetization data, i.e., the M(T) curve of 2.5% Eu-doped $Bi_2Se_3$ for an applied field of 1T. Interestingly, the M increases with a decrease in T, but no clear signature of long-range magnetic ordering is observed down to 2 K. Also, no evidence of hysteretic magnetic behavior was observed for the entire temperature range, as no bifurcation has been shown in the zero-field cooled (ZFC) and field-cooled (FC) cycles. This behavior is quite similar to the Dy-doped $Bi_2Se_3$.[70,71] The inset of Figure 6(a) shows a weak short-range magnetic anomaly near 50 K in the first derivative of the magnetic susceptibility d$\chi$/dT plot. The 5% Eu doped $Bi_2Se_3$ shows an antiferromagnetic ordering at 5.5 K.[72] May be with increasing in doping concentration causes more disordering and spin fluctuation that is observed in the present compound near 50 K. Figure 6(b) shows the isothermal magnetization at 10 K for 2.5% Eu-doped $Bi_2Se_3$. The magnetization shows non saturating behavior up to 2 T like Bi-rich $Bi_2Se_3$ nanoplates and Dy-doped $Bi_2Se_3$.[73,70]

To shed more light on the changing magnetic properties with Eu doping level, we have further studied the 10% Eu-doped $Bi_2Se_3$. Figure 6(c) shows the temperature dependence magnetization from 300K-2K for both the ZFC and FC cycle for the sample 10% Eu-doped $Bi_2Se_3$. Here, magnetization also increases with a decrease in temperature. No long-range magnetic ordering has been observed down to the lowest measured temperature, as in the case of the 2.5% doping one. The ZFC and FC cycles coincide with each other, as is expected for a typical paramagnetic sample. For a better understanding of the nature of the magnetic ground state of 10%, Eu doped $Bi_2Se_3$, the inverse dc susceptibility $\chi^{-1}$ ($\chi = M/H$) has been plotted as a function of temperature in the inset of Figure 6(c). At high temperatures above 170 K, the susceptibility follows the Curie-Weiss law, $\chi = C/(T - \theta_{CW})$, where C is the Curie constant and $\theta_{CW}$ is the Curie-Weiss temperature. From the linear fit to the high-temperature data, we have calculated the values of the effective paramagnetic moment $\mu_{eff} = 5.95\mu_B$/f.u. and $\theta_{CW} = -114$ K. A negative $\theta_{CW}$ suggests that the dominating exchange interaction in 10% Eu doped $Bi_2Se_3$ is antiferromagnetic. The above value of $\mu_{eff}$ is smaller than the effective moment of $Eu^{2+}$ (7.9$\mu_B$/Eu) ion in the PM state. This value is in between full magnetic moments of $Eu^{2+}$ (7.94 $\mu_B$) and $Eu^{3+}$ (3.4 $\mu_B$). A similar observation of the mixed magnetic state of Eu has been revealed in the previous report on Eu substitute $Bi_2Se_3$ single crystal.[72] With decreasing temperature, $\chi^{-1}$ starts to deviate from the linear behavior below ∼170 K, which is relatively high. This deviation of $\chi^{-1}$ (T) at high temperatures well above $\theta_{CW}$ may be due to the strong spin fluctuations in the PM state because of short-range magnetic correlations. To explore the influence of the magnetic field on the magnetic ground state, we have measured the field dependence of the magnetization (M-H) in 10% Eu doped $Bi_2Se_3$ up to 7 T at 10 K. A non-saturating magnetization has also been observed in this case as 2.5% Eu doped $Bi_2Se_3$.[71] No hysteresis in the magnetic field cycle has been detected. Our study on the doping concentration of Eu in $Bi_2Se_3$ has revealed that doping destroys the magnetic ordering of the system and enhances spin fluctuations.

## 3. Conclusions

Here, we have successfully synthesized $Bi_2Se_3$ and $(Eu_xBi_{1-x})_2Se_3$ MSs by solution-processable colloidal technique. The as-synthesized ultrathin MSs show hexagonal geometry as depicted from the TEM analysis. Here, PVP plays a vital role in the morphology of $Bi_2Se_3$, as it is used as a capping agent. Rare earth Eu doping in $Bi_2Se_3$ is established from the XRD pattern, which showed a clear peak shift with doping. The Raman spectra indicated that the basic lattice structure was retained in all the samples; however, the intensity profile was changed with doping concentration. The stoichiometric variation also significantly affected the optical properties of the material, which was reflected in the band gap reduction with doping. Enhancement in the band gap upon Eu doping is established from both STS measurement as well as theoretical studies. Magnetic studies show an antiferromagnetic ordering in the doped samples, which is unusual, and it can open up a new window toward future spintronic devices.

**Author contributions:**

**Sumana Paul:** Writing – original draft, Methodology, Investigation, Data collection, Formal analysis, Data curation, Conceptualization. **Moumita Das:** Writing, Formal analysis, Data collection and curation. **Sujoy Datta:** Writing, Theoretical calculation and analysis. **Raja Chakraborty:** Writing, Formal analysis, Data collection and curation. **Prabhat Mandal:** Formal analysis. **P. K. Giri:** Supervision, Formal analysis, Writing – review & editing.

**Conflicts of interest:**

There are no conflicts to declare.

**Acknowledgment:**

Author S.P. acknowledges the project no. NANO/IPDF/2021-22/SP/01 for providing the fellowship during the tenure of the work. P.K.G. acknowledge the financial support from SERB (Grant number CRG/2021/006397), MEITY (Grant No. 5(1)/2022-NANO), UGC-DAE CSR (Grant No. CRS/2021-22/01/349) for carrying out part of this work. Central Instruments Facility, IIT Guwahati, is acknowledged for providing the TEM, XRD, and Raman facilities. The instrumental facility of IACS is also acknowledged for providing TEM facility and absorbance.**Electronic Supplementary Information:** Supplementary material contains TEM, HRTEM, and mapping images of the undoped and Eu-doped $Bi_2Se_3$ NSs and high-resolution XPS spectra of doped $Bi_2Se_3$ MSs.

**REFERENCES**

1.      S. Kang, D. Lee, J. Kim, A. Capasso, H. S. Kang, J. Park, C. Lee, and G. Lee; 2D semiconducting materials for electronic and optoelectronic applications: potential and challenge, 2D Mater, 2020, 7, 022003.

2.      T. Tan, X. Jiang, C. Wang, B. Yao, and H. Zhang, 2D Material Optoelectronics for Information Functional Device Applications: Status and Challenge, Adv. Sci. 2020, 7, 2000058.

3.      M. Gibertini, M. Koperski, A. F. Morpurgo and K. S. Novoselov, Magnetic 2D materials and heterostructures, Nat. Nanotechnol., 2019, 14, 408–419.

4.      R. Mas-Balleste, C. Gomez-Navarro, J. Gomez-Herrero and F. Zamora, 2D materials: to graphene and beyond, Nanoscale, 2011, 3, 20–30.

5.      K. F. Mak and J. Shan, Photonics and optoelectronics of 2D semiconductor transition metal dichalcogenides, Nat. Photonics., 2016, 10, 216-226.

6.      L. Kou, Y. Ma, Z. Sun, T. Heine, and C. Chen, Two-Dimensional Topological Insulators: Progress and Prospects, J. Phys. Chem. Lett. 2017, 8, 1905−1919.

7.      O. Breunig and Y. Ando, Opportunities in topological insulator devices, Nat. Rev., 2022, 4, 184-193.
18

**FIGURES:**

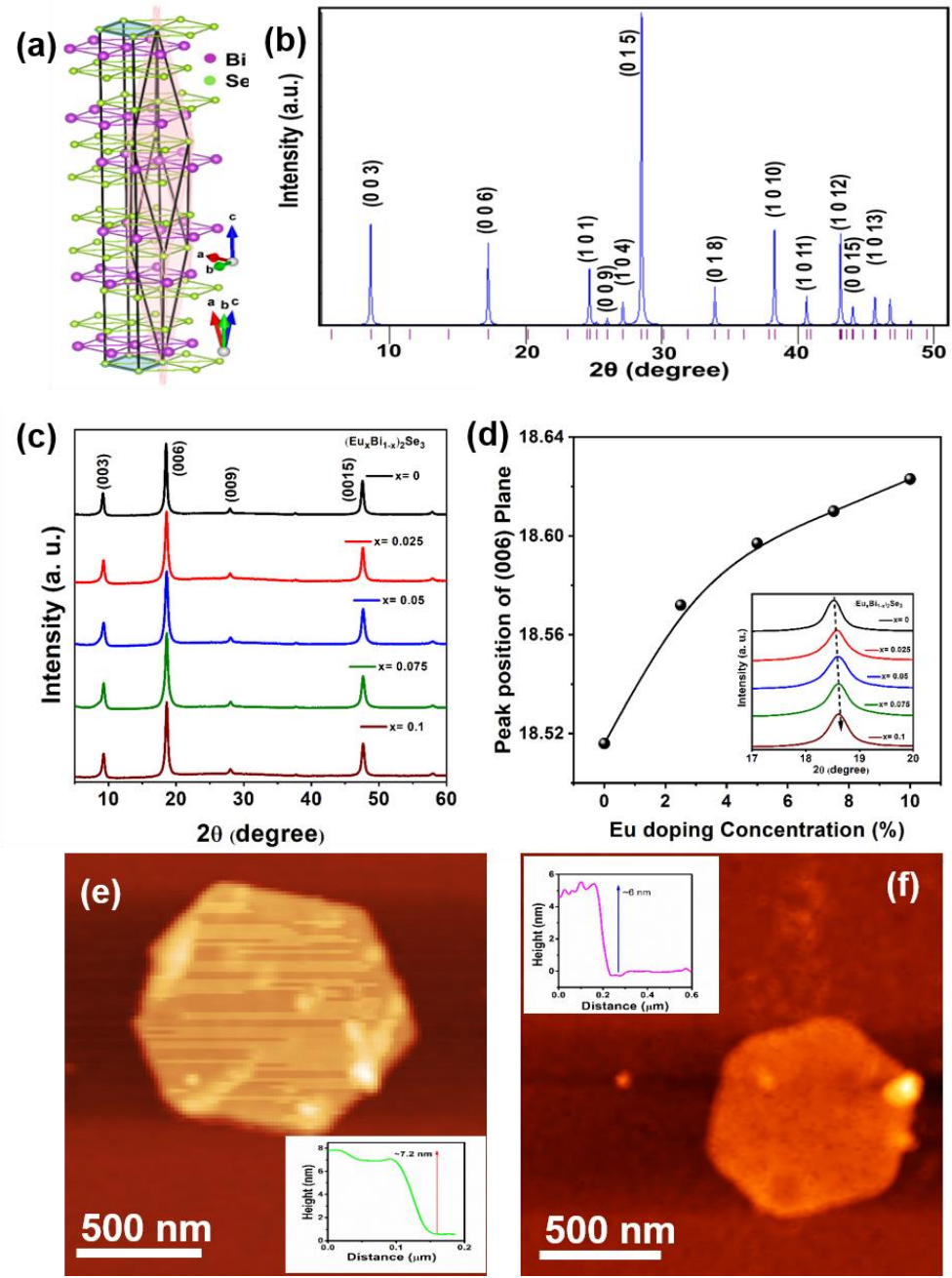

**Figure 1:** (a) Crystal structure of $Bi_2Se_3$, where the pink and blue shaded regions represent the primitive and conventional unit cells, respectively, (b) The simulated XRD pattern of the conventional unit cell. (c) XRD pattern of undoped and Eu-doped $Bi_2Se_3$ (($Eu_xBi_{1-x})_2Se_3$) nanosheets for x = 0, 0.025. 0.05, 0.075, 0.1. (d) Shift of the (003) peak position with Eu doping concentration, which follows a non-linear relationship. A zoomed-in view of the (006) XRD plane in the range 17°−20° which depicts a red shift upon Eu doping. AFM image of a single (e) undoped $Bi_2Se_3$ MS and (f) Eu-doped $Bi_2Se_3$ MS. The inset shows the height profile of MS on a Si substrate.

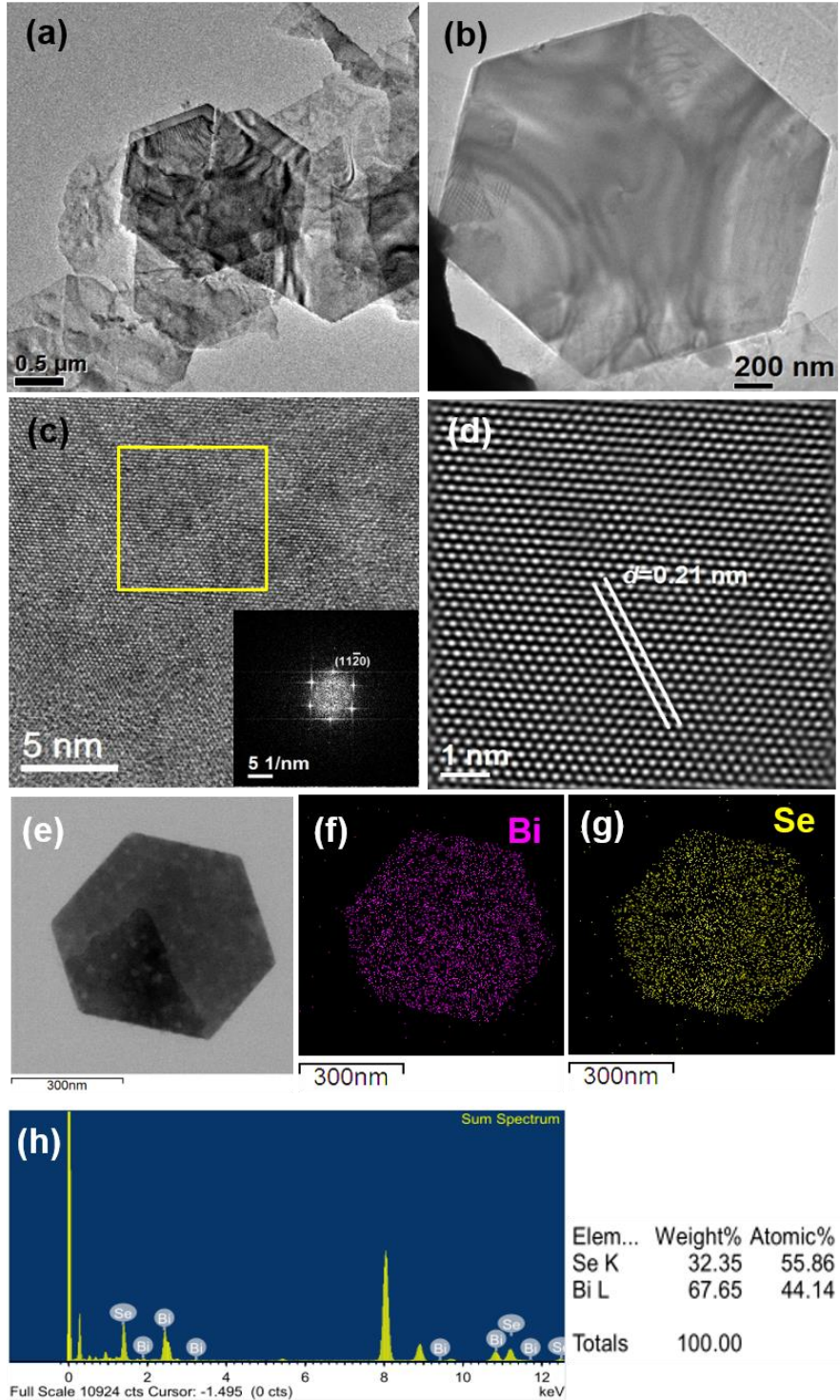

**Figure 2.** (a) Large area inverted TEM image of undoped $Bi_2Se_3$ MSs. (b) A closer view of a single MS. (c) HRTEM lattice image of a single MS; the inset shows the FFT pattern of the (11-20) group of planes as obtained from the yellow squared region of the HRTEM image of $Bi_2Se_3$. (d) Reconstructed HRTEM image showing the *d* spacing of 0.21 nm corresponding to the group of planes (11-20). (e) Scanning TEM image of $Bi_2Se_3$ nanosheet and corresponding STEM elemental mapping of (f) Bi, and (g) Se atoms, respectively. (h) EDX spectrum of $Bi_2Se_3$ MS showing the atomic percentage of different elements.

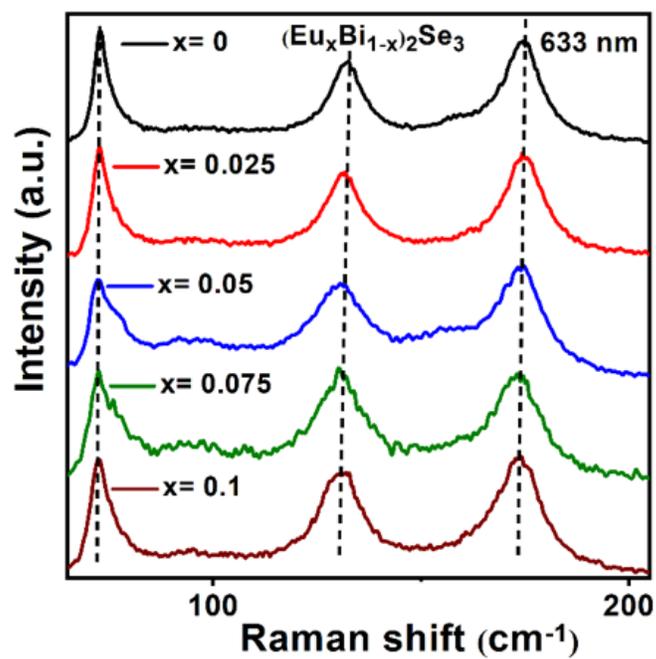

**Figure 3:** Raman spectra of undoped and Eu-doped $Bi_2Se_3$ (($Eu_xBi_{1-x})_2Se_3$) nanosheets for x = 0, 0.025. 0.05, 0.075, 0.1.

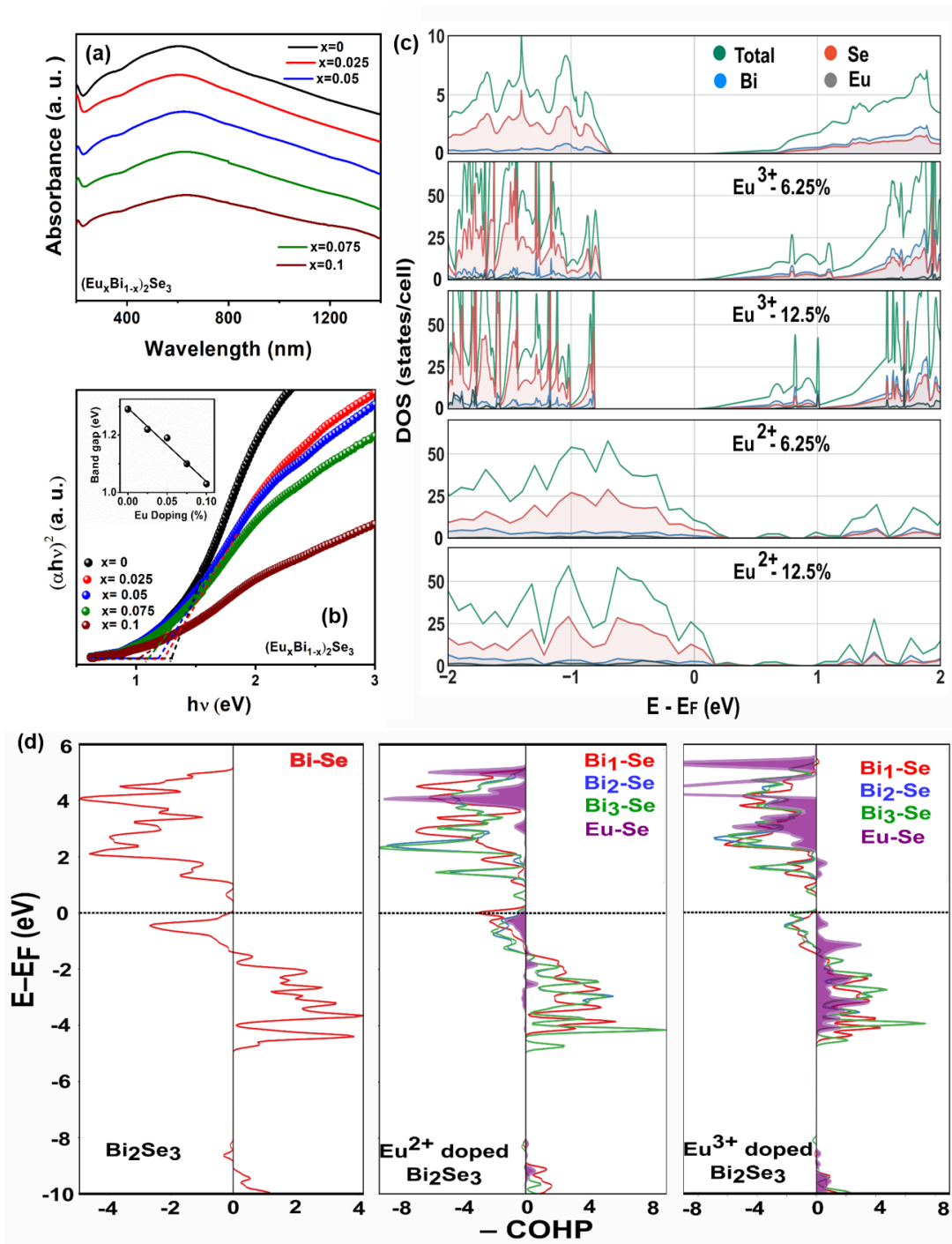

**Figure 4.** (a) Optical absorption spectra of undoped and Eu doped $Bi_2Se_3$ MSs in a wide spectral range (200-1400 nm). (b) Tauc plots for different samples showing variation of bandgap with Eu doping concentrations; the inset shows a linear variation (decrease) of bandgap with Eu doping concentration. (c) Electronic density of states of pristine and $Eu^{2+}$, $Eu^{3+}$ doped $Bi_2Se_3$ at different concentrations. (d) The negative Crystal orbital Hamilton population (-COHP) plot for pristine and doped Bi2Se3. Positive (-COHP) indicates bonding and negative (-COHP) indicates anti-bonding states.

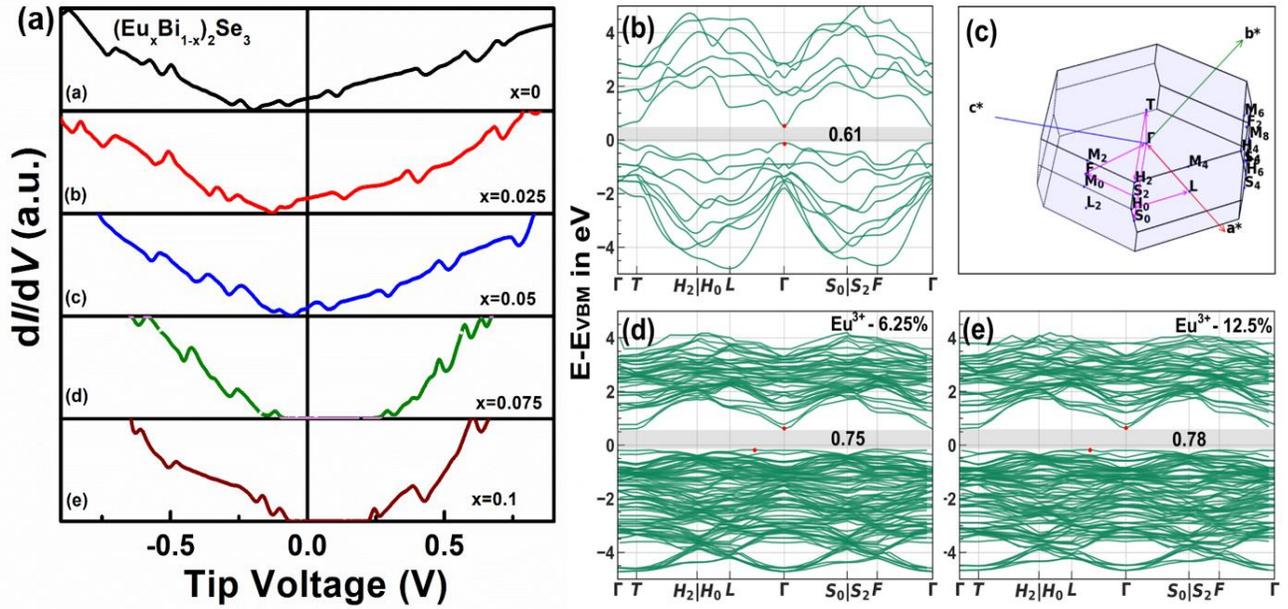

**Figure 5:** (a) dI/dV vs tip voltage of the CB and the VB energies with respect to Fermi energy (EF) from STS studies of undoped and Eu doped $Bi_2Se_3$ MSs. (b) Energy band structure of pristine bulk $Bi_2Se_3$; (c) the Brillouin Zone for the primitive cell of $Bi_2Se_3$; (d, e) band structure of $Eu^{3+}$ doped structure for 6.25% and 12.5% doping, respectively.

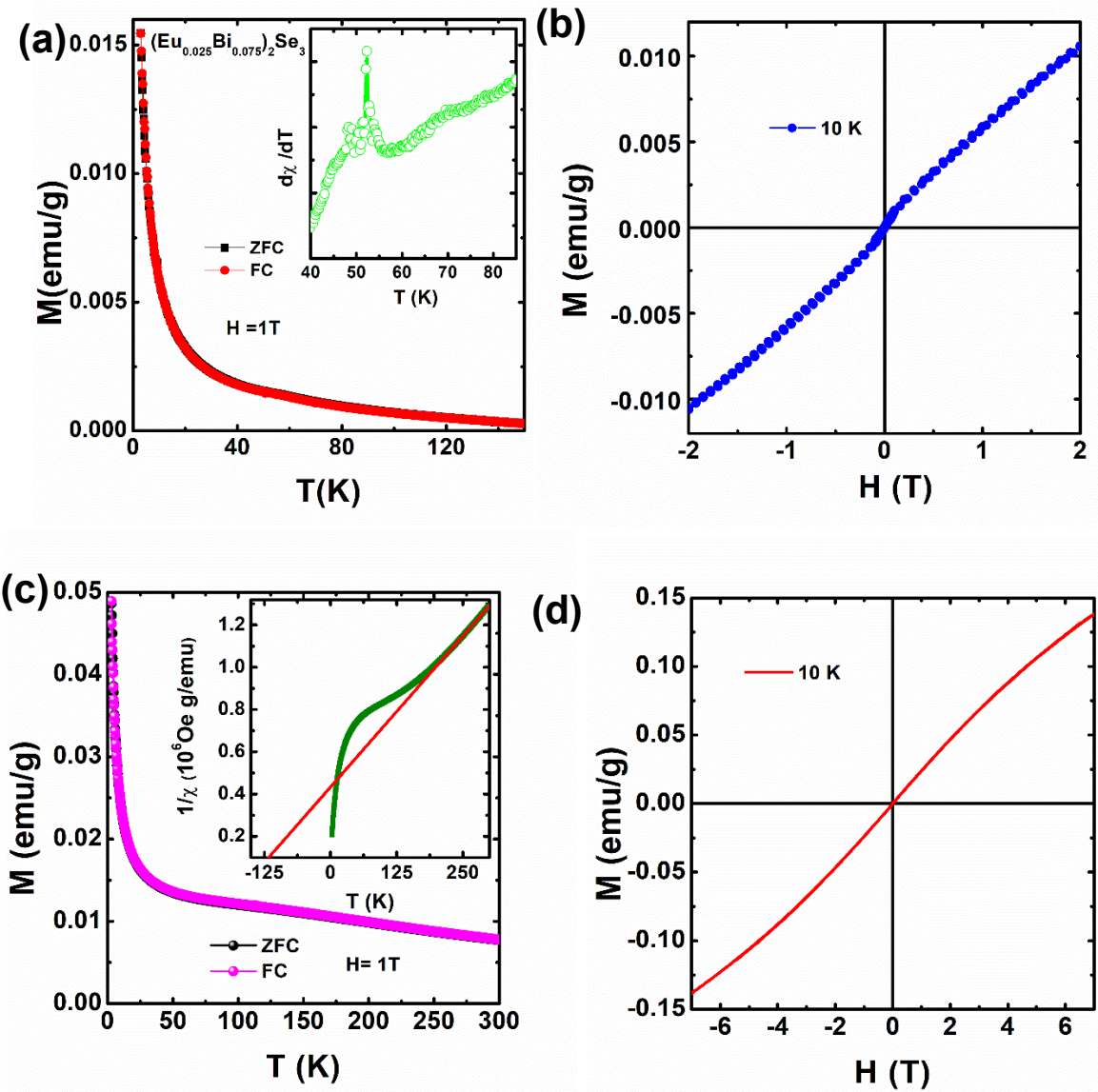

**Figure 6:** (a) Main panel shows the M(T) for both ZFC and FC cycle and inset shows the first derivative of susceptibility with T near 50 K. (b) shows the isothermal magnetization at 10 K for 2.5% Eu-doped $Bi_2Se_3$. (c) Main panel shows the M(T) for both ZFC and FC cycle and inset shows inverse susceptibility with T. (d) shows the isothermal magnetization at 10 K for 10% Eu-doped $Bi_2Se_3$.

TOC Image:

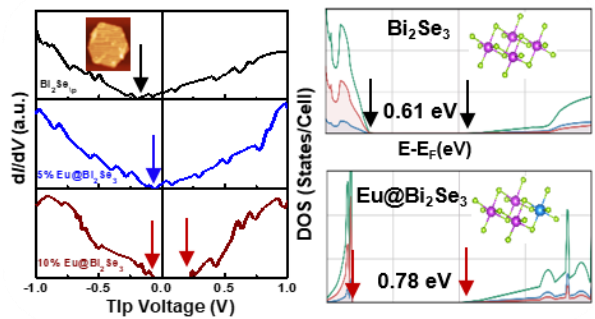